\documentclass{emulateapj}
\usepackage{graphicx}

\shorttitle{
Inner Rim of Molecular Disk around HD 141569 A
}
\shortauthors{Goto et al.} 
\begin{document}

\title{Inner Rim of A Molecular Disk Spatially Resolved in Infrared CO
  Emission Lines\altaffilmark{1}} 

\author{M. Goto,\altaffilmark{2} T. Usuda,\altaffilmark{3}
C. P. Dullemond,\altaffilmark{2}
Th. Henning,\altaffilmark{2}
H. Linz,\altaffilmark{2,4}
B. Stecklum,\altaffilmark{4}
\protect\and  H. Suto\altaffilmark{5}}

\email{mgoto@mpia.de}

\altaffiltext{1}{Based on data collected at Subaru Telescope, which is
                 operated by the National Astronomical Observatory of
                 Japan.}

\altaffiltext{2}{Max Planck Institute for Astronomy, 
                 K\"onigstuhl 17, D-69117 Heidelberg, Germany.}
     
\altaffiltext{3}{Subaru Telescope, 650, North A`ohoku Place, Hilo, HI
                 96720.}

\altaffiltext{4}{Th\"uringer Landessternwarte Tautenburg, Sternwarte 5,
                 D-07778 Tautenburg, Germany.}

\altaffiltext{5}{National Astronomical Observatory of Japan, Osawa,
                 Mitaka, Tokyo 181-8588, Japan.}

\begin{abstract}

We present high-resolution infrared spectroscopy of the Herbig Ae star
HD~141569~A in the CO~$v=2-1$ transition. With the angular resolution
attained by the adaptive optics system, the gas disk around
HD~141569~A is spatially resolved down to its inner-rim truncation.
The size of the inner clearing is 11$\pm$2~AU in radius, close to the
gravitational radius of the star. The rough coincidence to the
gravitational radius indicates that the viscous accretion working
together with the photoevaporation by the stellar radiation has
cleared the inner part of the disk.

\end{abstract}


\section{Introduction}

Direct imaging of circumstellar disks is one of the most rapidly
developing fields of observational astronomy (e.g. Grady et
al. 2005). It faces, however, two barriers in the optical/infrared: it
is mostly sensitive to the dust grains either in scattered light or
thermal emission, although dust carries only a fraction of the mass
with all the rest in the gas phase mostly in molecular form. Moreover,
it probes only outer regions of a disk, being obstructed by glaring
light from the central star.  However, it is this invisible part of
the disk, gas inside 30~AU of the star, that has direct consequence to
the gas giant planet formation. The gas dissipation at the inner disk
practically limits its time scale, and make a critical test to the
current planet formation theories that predict distinct time scales to
build up Jupiter-like planets ($\sim$10$^7$~yr, Pollack et al. 1996;
$\sim$10$^5$~yr, Boss 2002).

Molecular spectroscopy in the near-infrared has a particular advantage
in this area. The energy gap between the ground state and the
vibrationally excited states is on the order of 10$^3$~K, which
automatically guarantees that the gas in emission is in the innermost
part of the disks. The molecular spectroscopy also ensures that the
emission does not arise as close as in the stellar photosphere where
the temperature is too high to keep the molecules from dissociation.
The CO molecule is the versatile spectroscopic tracer of the gas disks
used by many authors both at the first overtone at
2.3~\micron~\citep{geb87,car89,cha93,naj96,bis97,kra00,bik04,blu04,thi05},
and the fundamental band at
4.7~\micron~\citep{bri03,naj03,bla04,ret04,car05}.

HD~141569~A is a Herbig Ae star in transition to a debris disk object
with an optically thin disk with relatively small infrared excess
($L_{\rm IR}/L_\ast = 8 \times 10^{-3}$ by Sylvester et al. 1996; cf.
$L_{\rm IR}/L_\ast = 2.4 \times 10^{-3}$ of $\beta$~Pic by Heinrichsen
et al. 1999). The dust disk around HD~141569~A has been directly
imaged with coronagraphy by {\it HST} in the scattered light at the
visible and the near-infrared wavelengths
\citep{aug99,wei99,mou01,cla03}, as well as a ground-based telescope
with adaptive optics system \citep{boc03}. The position angle of the
disk is 356\arcdeg$\pm$5\arcdeg~east from north in projection on the
sky \citep{wei99}. The disk ellipse in the thermal emission is well
aligned to the dust scattered light (355\arcdeg$\pm$19\arcdeg~by
Fisher et al. 2000; 354\arcdeg$\pm$4\arcdeg~by Marsh et al. 2002). The
residual gas in the outer disk has been detected in the radio
spectroscopy at CO~$J=1-0$ and $J=2-1$ ($M_{\rm H_2} =
300~M_{\earth}$, Zuckerman, Forveille, \& Kastner 1995; Dent, Greaves,
\& Coulson 2005). The line emission CO~$J=2-1$ is recently spatially
resolved by \citet{dut04} with red- and blue-shifted components split
in north and south of the star with a velocity interval indicative of
gas orbital motion.

HD~141569~A is exceptional among Herbig~Ae/Be stars for its CO~$v=2-1$
emission. The infrared transition $v=2-1$ manifests the molecules are
first electronically excited by the UV irradiation. \citet{bri03}
argues that the absence of high velocity wings in CO $v=2-1$ indicates
that the line emission comes from the inner-rim of the disk receded
from the star by 17~AU, where the disk wall is directly
illuminated. The presence of inner clearing of the size of 10--20~AU
is consistent with the lack of near-infrared excess in the infrared
energy distribution \citep{syl96}, and the mid-infrared imaging
implying disk within 30~AU is already cleared to some extent
\citep{mar02}. The size of the inner hole is well within the reach of
spatially resolved observations with adaptive optics systems at 8-m
class telescopes at the distance of the star ($d=$99--108~pc, van den
Ancker et al. 1998; Mer\'in et al. 2004; hereafter we take $d=$108~pc
from Mer\'in et al). The goal of this paper is to present such
directly resolved observations of this inner clearing in CO~$v=2-1$
transitions.

\section{Observation and Data Reduction}

The spectroscopic observation was carried out on 25 May 2005 UT at the
Subaru Telescope with the facility spectrograph IRCS
\citep{tok98,kob00}. The curvature sensing adaptive optics system was
used to feed nearly diffraction-limited images to the spectrograph
\citep{gae02,tak04}. HD~141569~A ($R$=7.02; Mer\'in et al. 2004)
itself was used as the wavefront reference source at the visible
wavelength. The grating angles were set so that the two echelle bands
at 4.597--4.718~$\mu$m and 5.014--5.147~$\mu$m were covered in the
1k$\times$1k InSb array. The first short-wavelength band includes the
$R$-branch of CO~$v=2-1$ at $R$(0) to $R$(15). The resolving power of
$R=$20,000 ($\Delta v=$15~km~s$^{-1}$) was attained using a narrow
slit of 0\farcs15$\times$7\farcs5.  The plate scale along the slit is
60~mas per pixel. The slit was oriented to north to south, parallel to
the major axis of the disk on the sky. The data were recorded by
switching the field of view by 3\arcsec~in every 5~minutes along the
slit with the tip-tilt mirror inside the adaptive optics system in
order to subtract the sky emission. The total integration time is
20~minutes. The spectroscopic flat field was obtained at the end of
the night from a halogen lamp exposure. The seeing was modest,
0\farcs9 at $R$, during the observation.

\begin{figure}
\includegraphics[angle=0,width=0.48\textwidth]{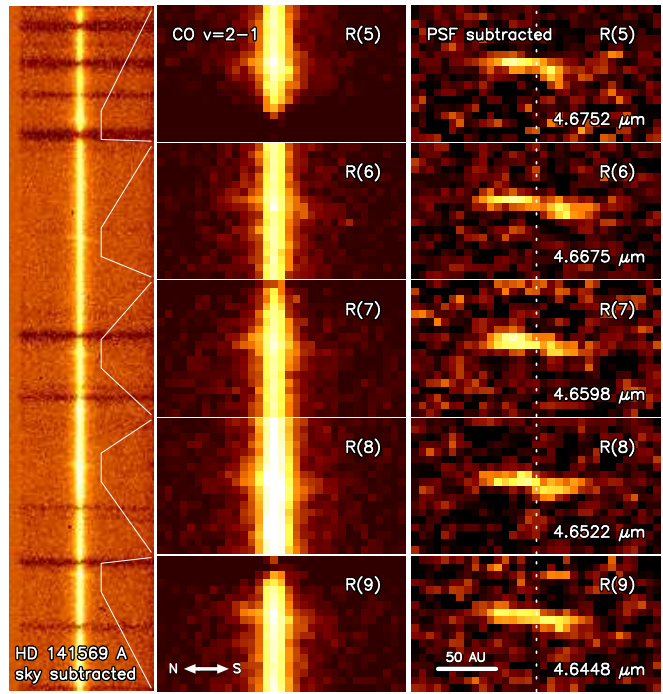}
\caption{Left: a cut-out of the spectrogram of HD~141569~A at
  4.6~\micron~obtained with IRCS at the Subaru Telescope on the night
  of 25th May 2005. The slit was oriented from north to south along
  the major axis of the disk projected on the sky, with north being
  the left hand side. The spatial and spectral distortion was
  corrected using emission lines in the telluric atmosphere. The
  emission lines of CO~$v=2-1$ within our spectral coverage are all
  found spatially extended up to 50 AU at both sides of the
  star. Middle column: the blow-up of the emission lines from $R$(5)
  to $R$(9). Right column: same as in the middle panels, but after
  one-dimensional point spread functions have been subtracted. The
  centrer of the PSF, therefore, the position of the star is marked by
  dotted verticals. Disk rotation is clearly visible, with the
  northern disk receding from us. The rotation is consistent with the
  spiral arms seen in the dust scattering \citep{cla03}, and the
  earlier radio spectroscopy (Dutrey et al. 2004; Figure~3 presented
  in their paper accidentally shows the northern component of the CO
  emission is approaching toward us. The confusion has been cleared by
  private communication with A. Dutrey and J.-C. Augereau).\label{f1}}
\end{figure}

Two-dimensional spectrograms were reduced in the standard manner with
the IRAF\footnote{IRAF is distributed by the National Optical
Astronomy Observatories, which are operated by the Association of
Universities for Research in Astronomy, Inc., under cooperative
agreement with the National Science Foundation.} image reduction
package, involving sky-subtraction, flat-fielding, and interpolation
of outlier pixels. The wavelength calibration was carried out by
maximizing the cross-correlation between the observed spectra and the
atmospheric transmission curve calculated by ATRAN \citep{lor92}. The
tilt of the dispersion axis with respect to the array column, and the
geometrical distortion, i.e., the curvature of the slit image on the
detector array, was corrected so that the dispersion and the telluric
emission lines show up straight to the detector vertical and
horizontal over whole spectral range. A part of the spectrogram after
the correction of spectral and spatial distortions is shown in
Figure~\ref{f1}.

The CO line emission at $v=2-1$ within our spectral coverage is all
found spatially extended up to 50~AU beyond the photospheric emission
of the star. In order to isolate the molecular lines from the stellar
photospheric emission, one-dimensional point spread function (PSF) was
taken from the nearby continuum, and was subtracted from the line
emission. Note that the spectral PSF, although spatially one
dimensional, provides almost an ideal PSF: it is recorded by the same
instrument at the same time at nearly identical wavelengths to the
line emission. The size of the PSF measured at the continuum
wavelength was 0\farcs24 in full width at half maximum. Caution was
taken at the scaling of the PSF, since over-subtraction of the PSF
introduces an artificial depression at the center, which is possibly
mistaken for a disk inner hole. First, one dimensional spectrum was
extracted inside a small aperture matched to the PSF to sample the
stellar continuum flux exclusively. The one dimensional spectrum was
smoothly interpolated from the both sides of a line to measure the
continuum contribution at the line wavelength. The PSF was scaled to
the height of the continuum flux, and subtracted from the line
emission accordingly (Figure~\ref{f1}).

We briefly discuss the influence of the telluric absorption below.
The interference of CO emission lines over the telluric absorption is
illustrated in the spectra in \citet{bri03} obtained with similar
spectral resolution with the present observation. The energy loss in
absorption in the telluric atmosphere has to be compensated to restore
the absolute flux in the line emission, and pottentially the source of
large systematic uncertainty in the flux measurement.  The radial
profile is, on the other hand, less subject to the atmospheric effect,
as long as the emission line stays at constant wavelength along the
radius. However, because of the finite displacement in the line
velocity by the gas orbital motion, the emission lines still have to
be at reasonably flat part of the transmission curve. Fortunately the
interference of the telluric absorption is less severe at CO~$v=2-1$
than CO~$v=1-0$ in which the telluric atmosphere has corresponding
absorption lines, as is also seen in \citet{bri03}. Since apparent gas
velocity at 5~AU and 20~AU is 15~km~s$^{-1}$ and 7~km~s$^{-1}$
respectively, if it is in the Keplerian rotation, a flat part of
$\Delta v\approx8$~km~s$^{-1}$ would suffice for the radial profile to
be intact. Those lines longword of 4.68~$\mu$m that seriously overlap
with the telluric absorption lines with the clearance less than this
interval were discarded, and the lines from $R$(5) to $R$(9) were only
used for the further analysis at reconstruction of the radial profile
and the line image shown in Figure~\ref{f2} and \ref{f3}.

\section{Result}

\begin{figure}
\includegraphics[angle=-90,width=0.48\textwidth]{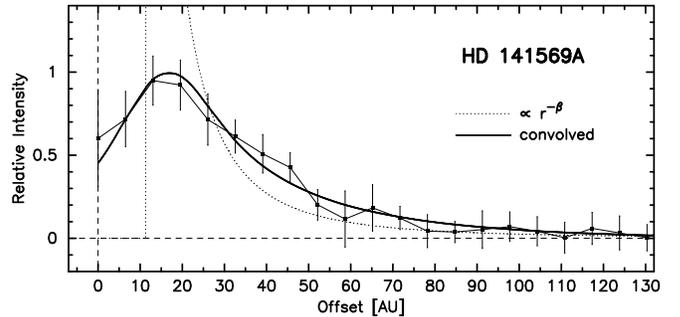}
\caption{The radial profile of CO line emission at the central part of
  the disk. A power-law radial profile (dotted line) was fit to the
  observation after convolved with the PSF (solid line) to find the
  best-fit cut-off radius at $r_i=11\pm 2$~AU. The error bars
  (1~$\sigma$) are given by the standard deviation of the radial
  profiles extracted from CO~$v=2-1$ $R$(5) to $R$(9).\label{f2}}
\end{figure}

\begin{figure}
\includegraphics[angle=0,width=0.48\textwidth]{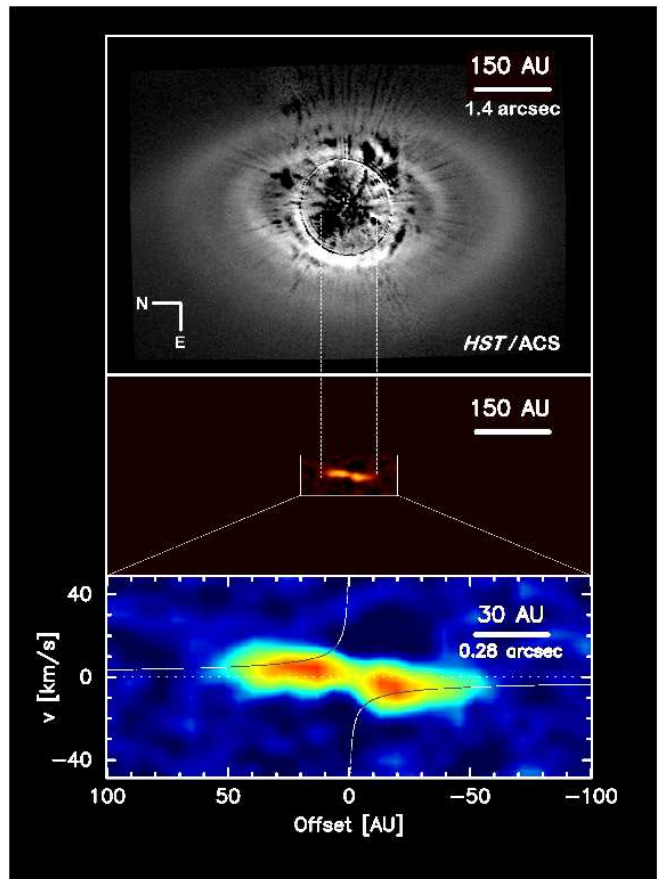}
\caption{Top: {\it HST}/ACS image of the debris disk around
  HD~141569~A from Clampin et al. 2003 at visible wavelength ($\lambda
  =$606~nm). The central 100~AU are masked by the coronagraph. Middle:
  the image of the composite line emission from CO~$v=2-1$
  $R$(5)--$R$(9). Only those lines less severly overlapped with the
  telluric absorption lines were used. The north is left, and the
  wavelength increases to the top. Bottom: the close-up view of the
  middle panel. The gap between the northern and the southern disks is
  clearly resolved. The velocity dispersion along the line of sight is
  calculated for a disk with an inner hole of 11~AU with the gas
  inside the disk in the Keplerian rotation (for assumed central mass
  of 2.0 $M_\odot$ with the inclination angle $\phi=$51\degr). The
  velocity contour overlaid in the shade is consistent with the
  observed line image. The rotation curve in the case of no truncation
  at the inner disk is shown in the dashed line. The lack of high
  velocity component close to the star represents that the molecular
  gas is cleared up already at the central part of the disk. There is
  a possible asymmetry seen in the brightness of the northern and
  southern disks, however, more observations are necessary to confirm
  it.\label{f3}}
\end{figure}

Each line image was summed up together along the dispersion axis by
$\pm19$~km~s$^{-1}$ to reconstruct a one-dimensional disk radial
profile. The central depression is clearly resolved in the combined
profile in Figure~\ref{f2}. Since the individual line image is
considered as an independent imaging, the error bars were given by the
standard deviation of the radial profiles at different transitions. A
power-law radial profile with an inner cut-off at $r_i$ ($f(r) \propto
r^{-\beta}$ at $r \geq r_i$, $f(r)=0$ at $r<r_i$) was fit to the
observation after convolved with the observed PSF. The best-fit radius
of the central clearing is found at $r_i=$11$\pm$2~AU with the
power-law index $\beta= $2.6$\pm$0.6. The combined disk image with
improved pixel sampling is shown in Figure~\ref{f3}. Disk rotation is
clearly detected as a velocity offset between the southern and
northern disks, approaching and receding from us,
respectively. HD~141569~A has its eastern disk near to us known from
the asymmetric azimuthal illumination of the disk, which is attributed
to the forward scattering of the grains \citep{wei99}. The disk
rotation is therefore clockwise in projection on the sky, which is
consistent with the winding of the spiral arms at 200--400~AU of the
star \citep{cla03}. A simple geometrical model is calculated for an
infinitely thin disk vignetted by the slit field-of-view with the gas
in the disk in the Keplerian rotation until the inner truncation
($M_\ast = 2.0 M_\odot$, Mar\'in et al. 2004; inclination angle $\phi
= 51$\arcdeg~from face-on, Weinberger et al. 1999). The velocity
dispersion along the line of sights well agrees with the observed disk
image in the line emission, and the lack of high velocity component
close to the central star manifests the innermost part of the disk is
indeed evacuated (Figure~\ref{f3}). The line emission from the
northern disk in Figure 3 looks slightly brighter than that from the
southern disk. The asymmetry might be also recognized in each line
profile shown in Figure 1 before it is combined together, however,
more observations are necessary to confirm the asymmetry, and discuss
further on the possible cause of it.

\section{Discussion}

The size of the central cavity gives some insights into its formation
mechanism. The inner clearing is apparently larger than the
magnetospheric truncation ($r\sim R_\ast \sim 10^{-2}$~AU), or the
dust sublimation radius ($r\sim 0.1$~AU), therefore those mechanisms
are not likely responsible. The cavity size is comparable to the
gravitational radius of the star where the sound speed in the ionized
medium is equal to the escape velocity of the system
($r_g=GM_\ast/c_i^2$; Shu et al. 1993, Hollenbach et al. 1994;
$\approx$18~AU for $M_\ast=2.0~M_\odot$).  The gravitational radius
defines the innermost radius of a disk where the gas is driven away
from the system once ionized by the stellar radiation. The rough
agreement of the inner edge of the disk around HD~141569~A with the
gravitational radius of the star indicates the cavity is being cleared
by the photoevaporation working in combination with the viscous
accretion.

The dissipation of an inner circumstellar disk is controlled by the
two processes: the mass loss by photoevaporation and the viscous
accretion on to the star. The viscous accretion is the primary drive
to remove a disk, transporting bulk of the material in the outer disk
inward eventually onto the star. On the other hand, the
photoevaporation is a slow process, steadily removing a disk outside
the gravitational radius as long as the extreme ultraviolet
irradiation (EUV; $h\nu>13.6$~eV) maintains. However, the persistency
or EUV radiation, or, the origin of the ionizing photon is unclear,
whether it is a consequence of active accretion \citep{muz01,muz04},
or of the stellar activity (e.g. Deleuil et al. 2005). In the former
case, the photoevaporation is expected to drop off as the viscous
accretion slows down, which causes problem to disperse a disk within
10$^7$~yrs \citep{mat03,rud04}. Although the photoionizing radiation
is likely chromospheric, and holds until the end of the disk evolution
in low-mass stars \citep{ale05}; it may not directly apply to
Herbig~Ae/Be stars \citep{gra05}. The mass loss via photoevaporation
proceeds most effectively immediate outside of $r_g$, as the disk wind
decreases with the radius as $\dot{\Sigma}(r) \propto r^{-5/2}$
\citep{hol94}. During the early phase of the disk evolution, the
photoevaporation is relatively unimportant, as the viscous accretion
quickly fills in the mass loss at the gravitational radius. However,
as the disk declines, the viscous accretion slows down, and is
eventually balanced by the photoevaporation. The net inflow across the
gravitational radius is quenched, as the photoevaporation virtually
reverses it. Starved for the further supply, the material inside $r_g$
is quickly drained onto the star on the time scale of viscous
evolution. In consequence, a disk dissipates on two distinct time
scales \citep{hol00,cla01,arm03,mat03,joh04,tak05,ale06}: the whole
disk lifetime may last as long as $\sim 10^7$~yrs, while the
evacuation of the inner disk is almost an instant ($\Delta t
\sim$10$^5$~yrs).

The presence of the central cavity clears several issues how the inner
disk is removed at the end of the disk dissipation. The
photoevaporation and the viscous accretion is likely effective to
remove a inner disk, inside outward, without calling for any
hypothetical planets to sweep it up. The radius of the inner clearing
measures $r_i=11$~AU, slightly smaller than the nominal gravitational
radius of $r_g=18$~AU. The smaller inner radius is in fact preferred
by the latest models that incorporate far ultraviolet radiation (FUV;
6~eV$<h\nu<$13.6~eV) that thermally heats up the gas in the disk
(Dullemond et al. 2006 for review). The mass loss by photoevaporation
formulated by \citet{shu93} and \citet{hol94} is actually the
photoionization that assumes dissipation of the gas ionized by
EUV. \citet{ada04} found FUV irradiation has significant contribution
to disk wind; and it could launch as close as 0.2--0.5~$r_g$ of the
central star, when the pressure gradient at the central region is
taken into account \citep{lif03,fon04}. The location of the inner-rim
of HD~141569~A better matches the new pictures.

HD~141569 is a triple system with low-mass companions B and C at
9\arcsec~away. The coeval formation of the system sets a robust age of
5$\pm$3 Myr \citep{wei00}, which agrees well with 4.7~Myr given by the
full spectral modeling with the photometric analysis of the primary
\citep{mar04}. The age of HD~141569~A sets a challenging time scale to
form gas-giant planets inside the gravitational radius of the
Herbig~Ae star. The conventional model of gas-giant formation takes
1--10~Myr to form a Jupiter-like planet by accretion of gas onto a
rocky core \citep{pol96,hub05}, which is uncomfortably close to the
standard disk lifetime ($\sim$3~Myr for the median disk lifetime,
Haisch et al. 2001; Hillenbrand 2005 for review), although many ideas
have been proposed to accelerate the core growth process
\citep{ina03,ali05,kle05,kla06}. The age of HD~141569~A suggests that
giant planets may not have enough time to fully deplete the
protoplanetary disk onto themselves, but likely have to compete with
disks in dissipation. The subtlety in the inner disk dissipation and
the planet forming time scale may explain a large scatter of disk
possession among the stars in a single age bracket (e.g. Rieke et
al. 2005), and eventually the diversity of the exoplanetary systems
discovered to date (e.g. Marcy et al. 2005).

It is tempting to speculate on the origin of the isotope anomaly in
the solar system in connection with the present observation. The
dissociation time scale of CO in the normal interstellar medium is on
the order of 10$^2$~yrs ($k_{\rm CO}\sim 2 \times 10^{-10}$~s$^{-1}$;
van Dishoeck 1988), the molecule in the disk of HD~141569~A will be
quickly destroyed without a shielding to the dissociating irradiation
from the star ($\lambda <$ 1100~\AA; van Dishoeck \& Black 1988). The
dust extinction could dilute the radiation \citep{kam00}, however, the
extinction cannot be too large, since the emission we observed here is
the transition from the vibrationally excited state ($v=2-1$), which
needs UV photons to get pumped up ($\lambda <$1600~\AA; Krotkov et
al. 1980). The margin is not too wide. The extinction dilemma could be
evaded by the line shielding either by molecular hydrogen, or by CO
itself \citep{van88}. In contrast to the dust extinction that dilutes
UV radiation as a continuum, line shielding blocks the dissociative
irradiation selectively. Self-shielding is only effective where the
dust extinction is less than $A_V=1$, but the CO is already optically
thick. It is not clear if any of these conditions are met in
HD~141569~A. The absence of near-infrared excess \citep{syl96} only
gives loose constraint, being consistent with the inner-rim as close
as 0.24~AU of the star \citep{mar04}; although the coronagraphic and
mid-infrared imaging suggests dust opacity drops at 30--160~AU and
inward \citep{wei99,mar02}. The column density of CO in the infrared
emission ($N_{\rm CO}\sim10^{11}$~cm$^{-2}$; Brittain et al. 2003) is
too short to make self-shielding effective \citep{lee96}, however, the
emitting CO does not necessarily properly represents the column
density of shielding gas along the line of sight to the UV source.

With all these uncertainties, if self line-shielding is still at work,
it may have an implication to the isotope anomaly imprinted in the
early solar nebula. The correlated isotope fractionation in $^{17}$O
and $^{18}$O in the primordial meteorites was once attributed to a
supernova outburst occurred near the solar system at its infancy
\citep{cla93}. Since then, numbers of accounts are proposed; injection
of nucleosynthesized material by evolved stars \citep{bus03}, direct
bombardment of high energy particles \citep{jal05}, and isotope
fractionation by self-shielding of CO \citep{cla02}. When the
self-shielding operates, CO is fractionated to C$^{16}$O as the rare
isotopomers C$^{17}$O and C$^{18}$O are readily destroyed for their
lower opacity, leaving more $^{17}$O and $^{18}$O in atomic form. The
isotope enrichment in the atomic gas potentially explains the
anomalous isotope ratio in the present solar system
\citep{yur04,lyo05}. Our observation found CO~$v=2-1$ emission at the
inner-edge of the disk from the star at the distance where planets
presumably form.  We might be witnessing the ongoing isotope
fractionation similar to that might happen in the early history of our
solar system.

\acknowledgments

{We thank all the staff and crew of Subaru for their assistance in
obtaining the data. We thank Diethard Peter for careful reading of the
draft. We thank Anne Dutrey, Jean-Charles Augereau, and Inga Kamp for
stimulating discussion. We thank Mark Clampin for his kind allowance
to reproduce his ACS image in Figure 3. We thank the anonymous referee
for the constructive suggestions that has critically improved the
present paper. M.G. is supported by a Japan Society for the Promotion
of Science fellowship.}

\end{document}